\documentclass[11pt]{article}

\usepackage[utf8]{inputenc}
\usepackage{geometry}
\usepackage{amsmath,amssymb}
\usepackage{mathrsfs}
\usepackage{natbib}
\usepackage{graphicx}

\newtheorem{remark}{Remark}

\renewcommand{\mathcal}{\mathscr}
\renewcommand{\d}[1]{\ensuremath{\operatorname{d}\!{#1}}}
\newcommand{\D}[1]{\ensuremath{\operatorname{D}\!{#1}}}
\newcommand{\dt}[1]{\frac{\d{#1}}{\d{t}}}
\newcommand{\Dt}[1]{\frac{\D{#1}}{\D{t}}}
\newcommand{\del}[2]{\frac{\delta{#1}}{\delta{#2}}}
\def\const{\ensuremath{\operatorname{const}}}

\def\cp{\ensuremath{\operatorname{\circlearrowleft}}}

\def\dim{\ensuremath{\operatorname{dim}}}
\def\Id{\ensuremath{\operatorname{Id}}}
\def\curl{\nabla\times}

\def\O{\boldsymbol\Omega}
\def\S{\boldsymbol\Sigma}
\def\A{\mathbf M}
\def\I{\mathbb I}

\def\a{\mathbf a}
\def\b{\mathbf b}
\def\c{\mathbf c}
\def\x{\mathbf x}
\def\f{\mathbf f}
\def\u{\bar{\mathbf u}}
\def\v{\mathbf v}
\def\m{\bar{\mathbf m}}
\def\t{\bar\vartheta}
\renewcommand{\tt}{\bar\theta}
\def\p{\varphi}

\def\z{\hat{\mathbf z}}
\def\n{\hat{\mathbf n}}

\def\J{\mathbb J}
\def\L{\mathcal L}
\def\H{\mathcal H}

\def\C{\mathcal C}
\def\U{\mathcal U}
\def\V{\mathcal V}
\def\W{\mathcal W}

\def\Fm{\U_{\m}}
\def\Gm{\V_{\m}}
\def\Hm{\W_{\m}}
\def\Fs{\U_{\rho_\alpha}}
\def\Gs{\V_{\rho_\alpha}}
\def\Hs{\W_{\rho_\alpha}}

\begin{document}

\title{Geometry of shallow-water dynamics with thermodynamics}

\author{F.J.\ Beron-Vera\\ Department of Atmospheric Sciences\\
Rosenstiel School of Marine \& Atmospheric Science\\ University of
Miami\\ Miami, FL 33145}

\maketitle

\begin{abstract}
  We review the geometric structure of the IL$^0$PE model, a rotating
  shallow-water model with variable buoyancy, thus sometimes called
  ``thermal'' shallow-water model.  We start by discussing the
  Euler--Poincar\'e equations for rigid body dynamics and the
  generalized Hamiltonian structure of the system.  We then reveal
  similar geometric structure for the IL$^0$PE.  We show, in
  particular, that the model equations and its (Lie--Poisson)
  Hamiltonian structure can be deduced from Morrison and Greene's
  (1980) system upon ignoring the magnetic field ($\vec{\mathrm B}
  = 0$) and setting $U(\rho,s) = \frac{1}{2}\rho s$, where $\rho$
  is mass density and $s$ is entropy per unit mass.  These variables
  play the role of layer thickness ($h$) and buoyancy ($\t$) in the
  IL$^0$PE, respectively. Included in an appendix is an explicit
  proof of the Jacobi identity satisfied by the Poisson bracket of
  the system.
\end{abstract}

\tableofcontents

\section{Introduction}

\begin{figure}[t!]
  \centering%
  \includegraphics[width=.5\textwidth]{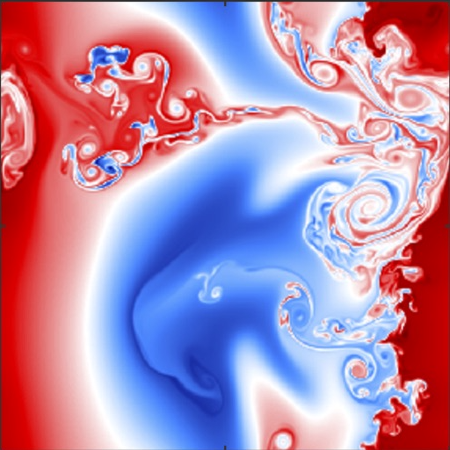}%
  \caption{Snapshot of potential vorticity from a numerical solution
  of a quasigeostrophic version of the IL$^0$PE model
  \eqref{eq:mom}--\eqref{eq:buo} in a doubly periodic domain
  $[0,1]\times [0,1]$.  Note the Kelvin--Helmholtz-like rollup
  filaments (length is scaled by the deformation radius of the
  system).}
  \label{fig:il0}%
\end{figure}

Following notation introduced by Ripa,\citep{Ripa-JFM-95} IL$^0$PE
stands for \emph{inhomogeneous-layer primitive-equation(s)} with
the superscript indicating that buoyancy does \emph{not} vary in
the vertical, while it is allowed to unrestrainedly vary in horizontal
position and time.  The IL$^0$PE has the \emph{two-dimensional}
structure of a \emph{rotating shallow-water model} with an additional
evolution equation for the buoyancy. This model was extensively
used through the 1980s and 1990s \citep{McCreary-85, Schopf-Cane-83,
McCreary-Kundu-89, Beier-97} to investigate mixed-layer (upper
ocean) dynamics as it allows one to accommodate, in a two-dimensional
setting and thus more easily, heat and freshwater fluxes through
the ocean's surface.  Such ``thermal shallow-water'' modeling was
abandoned due in large part to the increase of computational power
and a preference---I dare to say---to emphasize reproducing
observations over understanding the basic aspects of the dynamics.
However, a gratifying surprise has been to learn that this type of
modelling is regaining momentum,\citep{Zeitlin-18, Holm-etal-21}
particularly for the ability of the model to produce small-scale
circulations similar to those observed in ocean color images, even
at low frequency (Fig.\ \ref{fig:il0}). This renewed interest
motivated investigating the geometric properties of the system
further,\citep{Beron-21-PoF, Beron-21-RMF} following pioneering
work by Ripa.\citep{Ripa-GAFD-93, Ripa-JFM-95, Ripa-JGR-96,
Ripa-DAo-99} We review those properties here and also establish an
explicit connection, so far overlooked, with seminal work by
\citet{Morrison-Greene-80a} on generalized Hamiltonians.  The
exposition starts by reviewing similar geometric structure for
rigid-body dynamics.  We also include an explicit proof in an
appendix of the Jacobi identity that the Poisson bracket of the
model equations must satisfy.

\section{Background: The rigid body}

The free rigid body (Euler) equations in principle axis coordinates
are
\begin{equation}
  \I\dot\O = \I\O\times \O
  \label{eq:rb-ep}
\end{equation}
where $\I$ is the (diagonal) tensor of inertia and $\O(t)$ is
the angular velocity of the body. 

\subsection{Euler--Poincar\'e equations}

These equations follow from the variational principle
\begin{equation}
  \delta\int_{t_0}^{t_1}L(\O)\d{t} = 0,\quad 
  L(\O) := \tfrac{1}{2}\I\O\cdot\O
\end{equation}
with \emph{constrained} variations
\begin{equation}
  \delta\O = \dot\S + \S\times\O
\end{equation}
for some vector $\S(t)$ such that it vanishes at the endpoints. The
function $L(\O)$ is the \emph{Lagrangian}. The resulting equations
\begin{equation}
  \dt{}{}\frac{\partial L}{\partial \O} = \frac{\partial
  L}{\partial \O} \times \O
\end{equation}
are known as the \emph{Euler--Poincar\'e equations}.\citep{Marsden-Ratiu-99}

\subsection{Generalized Hamiltonian structure}

Using the angular momentum $\A := \I\O$, system \eqref{eq:rb-ep}
reads
\begin{equation}
  \dot\A = \A\times \I^{-1}\A.
  \label{eq:rb-lp}
\end{equation}
This set can be obtained by the \emph{Legendre transform}
$\O\mapsto\A$ defined by
\begin{equation}
  H(\A) = \A\cdot\O - L(\O)
\end{equation}
and the \emph{Poisson tensor}\footnote{The Poisson tensor is variously
called \emph{cosymplectic matrix} or, perhaps most appropriately,
\emph{Hamiltonian bivector}.} $\J$ with components
\begin{equation}
  \J^{ij} = \varepsilon^{ij}_k M^k 
\end{equation}
according to
\begin{equation}
  \dot\A = \J\nabla H(\A),
\end{equation}
which provides a \emph{generalized Hamiltonian} \citep{Salmon-88a,
Shepherd-90, Ripa-RMF-92, Morrison-98} formulation for the rigid body.  The
function $H(\A)$ is the \emph{Hamiltonian}. Associated with $\J$
is the \emph{Poisson bracket}, defined and given by
\begin{equation}
  \{U,V\} := \nabla U(\A)\cdot \J\nabla V(\A) = - \A \cdot \nabla
  U(\A)\times \nabla V(\A).
\end{equation}
This bracket is of the \emph{Lie--Poisson} type, i.e, linear in the
phase space coordinate $\A$, and (thus) satisfies 
\begin{align}
  & \{U,V\} = - \{V,U\}\quad \text{(antisymmetry)},
  & \{U,\{V,W\}\} + \cp = 0\quad \text{(Jacobi identity)}.
\end{align}

\subsubsection{Casimirs}

The Hamiltonian (energy) is an \emph{integral of motion}, clearly
since $\dot H = \{H,H\} \equiv 0$.  But the dynamics are constrained
by additional \emph{conservation laws}.  More precisely, because
$\det\J = 0$, i.e., $\J$ is singular, there exist functions $C(\A)$,
called \emph{Casimirs} (Lie’s  distinguished functions), whose
gradients span the \emph{null space} of $\J$, namely, they satisfy
\begin{equation}
  \J\nabla C(\A) = 0.
\end{equation}
These are given by
\begin{equation}
  C(\A) = F(\tfrac{1}{2}|\A|^2)\quad \forall F(\,).
\end{equation}
Note that $C(\A)$ \emph{commutes} with any function of state in
the Poisson bracket, viz.,
\begin{equation}
  \{U,C\} = 0\quad \forall U(\A);
\end{equation}
hence, they are conserved under the dynamics: $\dot C = \{C,H\}
\equiv 0$.  Note that the equations of motion are not altered
under the change of Hamiltonian $H \mapsto H + \lambda C$,
$\lambda = \const$.  The extremal points of $H$, however, may
be altered under this change. 

\subsubsection{Geometry}

More generally, let $z(t)$ represent a point in a space $M$ equipped
with a Poisson bracket $\{\,,\hspace{.1em}\}$.  One calls the pair
$(M,\{\,,\hspace{.1em}\})$ a Poisson manifold. The dynamical system
$\dot z = \J\nabla H(z) = \{z,H\}$ represents a (generalized)
Hamiltonian system.  More broadly, $\dot F = \{F,H\}$ for any
function of state $F(z)$. If the Poisson tensor (matrix) is singular,
then $\dim M = m$ must necessarily be odd.  This is different than
\emph{canonical} Hamiltonian dynamics for which $\dim M = 2n$.
Indeed, in such a case $z = (q,p)$, which satisfy $\dot q = +
\partial_p H$ and $\dot p = - \partial_q H$.  Thus
\begin{equation}
  \J = 
  \begin{pmatrix}
	 0 & +\Id^{n\times n}\\
    -\Id^{n\times n} & 0
  \end{pmatrix},\quad \det\J = 1,
\end{equation}
which is called \emph{symplectic matrix}.  It turns out that an
$m$-dimensional manifold is generically foliated by $2n$-dimensional
surfaces $\{C = \const\}$, called \emph{symplectic  leaves}, on
which the dynamics is canonical (clearly, if $z(0)$ lies on $\{C =
\const\}$, $z(t)$ will remain there for all $t$).

\subsubsection{Noether's theorem}

Finally, Noether's theorem relates \emph{symmetries} with
\emph{conservation laws}.  Energy is related with symmetry under
$t$ shifts and $s$-momentum with $s$-translational symmetry.  Casimirs
are not associated with explicit symmetries, but rather with
symmetries lost in the process of reducing a canonical Hamiltonian
system to a generalized (i.e., singular) Hamiltonian system. For
instance, a canonical system with dimension, say, $2n = 4$, that
has one integral of motion $I$ (say) can be reduced to a singular
system  with dimension $m = 3$ when it is constrained to the manifold
$\{I = \const\}$.  Such a loss of explicit symmetries happens in
fluid systems when formulated in Eulerian variables: the Casimirs
of hydrodynamics are related to the symmetry of the Eulerian variables
under Lagrangian particle relabelling,\citep{Newcomb-67, Ripa-AIP-81,
Padhye-Morrison-96} yet with a possibly important
caveat.\citep{Charron-Zadra-18}

\section{The IL$^0$PE model}

The IL$^0$PE model in some closed domain $D$ of the $\beta$
plane in a \emph{reduced-gravity} setting is given by
(e.g., Ripa \citep{Ripa-JGR-96})
\begin{align}
  \partial_t\u + (\u\cdot\nabla)\u + f\z\times\u +
  \tfrac{1}{2}h^{-1}\nabla h^2\t &= 0,\label{eq:mom}\\
	\partial_t h + \nabla\cdot h\u &= 0,\label{eq:vol}\\
	\partial_t\t + \u\cdot\nabla\t &= 0,\label{eq:buo}
\end{align}
where velocity $(\u)$, layer thickness ($h$) and buoyancy ($\t$)
are functions of horizontal position $(\x)$ and time ($t$).
Appropriate boundary conditions are
\begin{equation}
  \u\cdot\n = 0,\quad \n\times\nabla\t = 0\quad (\x\in\partial
  D).
\end{equation}
The condition on the left simply means no flow through the boundary
of $D$; the condition on the right (viz., that the boundary is
isopycnic) is needed to convey the IL$^0$PE a generalized Hamiltonian
structure.

\begin{remark}
The parenthesis in $(\a\cdot\nabla)\b$ is not superfluous!  Indeed,
\begin{equation}
  \a\cdot\nabla\b = \a^\top\nabla\b = (\nabla\b)^\top\a =
  (\nabla\b)\cdot\a = a_j\nabla b_j 
\end{equation}
while
\begin{equation} 
  (\a\cdot\nabla)\b = (\nabla\b)\a = a_j\partial_j\b
\end{equation}
(assuming that vectors are column and $\nabla\a$ is a matrix with
rows given by $\nabla a_i$).
\label{rem:vector}
\end{remark}

\begin{remark}
The set $\{\t = g' = \const\}$ is an invariant manifold of the
IL$^0$PE on which the dynamics are controlled by the HLPE, viz.,
$\partial_t\u + (\u\cdot\nabla)\u + f\z\times\u + g'\nabla h = 0$
and $\partial_t h + \nabla\cdot h\u = 0$.
\end{remark}

For the purpose of revealing the geometric structure of the IL$^0$PE,
it is convenient to write the momentum equation \eqref{eq:mom} in
two different but equivalent forms:
\begin{equation}
  \Dt{}(\u+\f) + (\nabla\u)\cdot(\u+\f) + \nabla\big(h\t -
  \tfrac{1}{2}|\u|^2 - \u\cdot\f\big) - \tfrac{1}{2}h\nabla\t = 0
  \label{eq:ep}
\end{equation}
and\footnote{Indeed, $\nabla\tt - \tfrac{1}{2}h\nabla h^{-1}\tt =
\tfrac{1}{2}\nabla h\tt$.}\footnote{Note that $(\u\cdot\nabla)\m  +
\m(\nabla\cdot\u) + (\nabla\u)\cdot\m = \bar m_j\nabla\bar u_j +
\partial_j \m\bar u_j$.}
\begin{equation}
  \partial_t\m +
  \underbrace{(\u\cdot\nabla)\m}_{\text{advection}}  +
  \underbrace{\m(\nabla\cdot\u)}_{\text{expansion}}  +
  \underbrace{(\nabla\u)\cdot\m}_{\text{stretching}} +
  h\nabla\big(\tfrac{1}{2}\tt - \tfrac{1}{2}|\u|^2 - \u\cdot\f\big) +
  \tfrac{1}{2}\tt\nabla h = 0
  \label{eq:lp}
\end{equation}
where
\begin{equation}
  \Dt{} := \partial_t(\cdot) + (\u\cdot\nabla)(\cdot),\quad 
  \curl\f := f\z,\quad 
  \m := h(\u+\f),\quad
  \tt := h\t.
\end{equation}
Here $\f$ is a vector potential for (twice) the local angular
velocity of the planet, and $\m$ is the momentum density dual
(conjugate) to $\u$ (cf.\ below). In particular, $\m\cdot\hat{\mathbf
x}$ with $\f = - (f_0 y + \tfrac{1}{2}\beta y^2)\hat{\mathbf x}$
gives the absolute angular momentum density (with respect to the
center of the planet and in the direction of the axis of rotation)
with an error of the order of the inverse of the planet's
radius.\citep{Ripa-JPO-97b} Note that $\tt$ represents a \emph{density
(form)}, rather than an advected quantity, just as $h$ (upon invoking
volume conservation). In getting \eqref{eq:ep} the following
\emph{fundamental} vector identity was used:
\begin{equation}
  \boxed{
  (\a\cdot\nabla)\b = (\curl \b)\times \a + (\nabla\b)\cdot\a,
  }
  \label{eq:vector}
\end{equation}
\emph{which can be written in several ways} (Rem.\
\ref{rem:vector}).\footnote{Indeed, $\Dt{}\u + (\curl\f) \times\u
= \Dt{}(\u + \f) + (\nabla\u)\cdot (\u+\f) - (\nabla\u)\cdot\f -
(\u\cdot\nabla)\f + (\curl\f) \times\u - \nabla\tfrac{1}{2}|\u|^2
= \Dt{}(\u + \f) + (\nabla\u)\cdot (\u+\f) + (\nabla\f)\cdot\u -
(\u\cdot\nabla)\f + (\curl\f) \times\u - \nabla(\tfrac{1}{2}|\u|^2
+ \u\cdot\f) =  \Dt{}(\u + \f) + (\nabla\u)\cdot (\u+\f) -
\nabla(\tfrac{1}{2}|\u|^2 + \u\cdot\f).$} This version uses ``mixed''
notation, to avoid confusion.  A consistent, but potentially
confusing, version would be $(\a\cdot\nabla)\b = (\curl \b)\times
\a + \a\cdot\nabla\b$.  Sometimes writing the last term using index
summation is useful.  Note, in particular, that $\smash{(\nabla\u)\cdot\u
= \nabla\frac{1}{2}|\u|^2}$. Equation \eqref{eq:lp} followed by
multiplying \eqref{eq:ep} by $h$ and using volume conservation
\eqref{eq:vol}. The components of $(\u\cdot\nabla)\m + \m(\nabla\cdot\u)
+ (\nabla\u)\cdot\m$ are $\bar m_j\partial \bar u_j + \partial_j
\bar m_i\bar u_j$.

\subsection{Euler--Poincar\'e variational formulation}

Consider the variational principle
\begin{equation}
  \delta\int_{t_0}^{t_1}\L[\u,h,\t]\d{t} = 0
\end{equation}
with \emph{constrained} variations
\begin{equation}
  \delta\u = \partial_t\v - (\u\cdot\nabla)\v + (\v\cdot\nabla)\u,\quad 
  \delta h = - \nabla\cdot h\v,\quad
  \delta\t = - \v\cdot\nabla\t,
\end{equation}
where $\v(\x,t_0) = 0 = \v(\x,t_1)$ is arbitrary. Its solution,
known as the \emph{Euler--Poinca\'re} equations,\citep{Holm-etal-02}
is given by\footnote{In reality, equation \eqref{eq:EP} follows
most directly in the form \eqref{eq:lp} prior to using volume
conservation, viz.,
\begin{equation}
  \partial_t\del{\L}{\u} + (\u\cdot\nabla)\del{\L}{\u}  +
  \del{\L}{\u}(\nabla\cdot\u) +
  (\nabla\u)\cdot\del{\L}{\u} - h\nabla\del{\L}{h} +
  \del{\L}{\t}\nabla\t = 0.
\end{equation}
It's just that the Kelvin circulation (Sec.\ \ref{sec:KN}) follows
directly using \eqref{eq:EP}, and thus is more convenient.}
\begin{align}
  \Dt{}\frac{1}{h}\del{\L}{\u} + (\nabla\u)\cdot
  \frac{1}{h}\del{\L}{\u}  - \nabla\del{\L}{h} +
  \frac{1}{h}\del{\L}{\t}\nabla\t &= 0,\label{eq:EP}\\
  \partial_t h + \nabla\cdot h\u &= 0,\\
  \Dt{\t} &= 0.
\end{align}

The Lagrangian \citep{Holm-etal-21}
\begin{equation}
  \L[\u,h,\t] := \int_D \tfrac{1}{2}h|\u|^2 + h\u\cdot\f -
  \tfrac{1}{2}h\t^2
  \label{eq:L}
\end{equation}
($\int_D$ acts on anything on the right), with variational
derivatives
\begin{equation}
  \del{\L}{\u} = h(\u+\f),\quad
  \del{\L}{h} =  \tfrac{1}{2}|\u|^2 = \u\cdot\f - \t h,\quad
  \del{\L}{\t} = - \tfrac{1}{2}h^2,
\end{equation}
gives the IL$^0$PE (with the momentum equation written as
\eqref{eq:ep}, most directly).  

\subsubsection{Kelvin circulation theorem}\label{sec:KN}

Let $D(\u) \subset D$ be a material region, i.e., \emph{transported}
by the flow of $\u$. Defining the \emph{circulation}
\begin{equation}
  \mathcal I := \oint_{\partial D(\u)}\frac{1}{h}\del{\L}{\u}\cdot\d{\x}
  \label{eq:I}
\end{equation}
from \eqref{eq:EP} it follows that
\begin{equation}
  \dt{\mathcal I} = \oint_{\partial D(\u)} \Dt{}\left(\frac{1}{h}\del{\L}{\u}\right)
  \cdot\d{\x} + \frac{1}{h}\del{\L}{\bar u_j}\nabla \bar u_j\cdot\d{\x} = -
  \int_{ D(\u)}
  \left[\frac{1}{h}\del{\L}{\t},\t\right]\d{}^2\x,
  \label{eq:KN}
\end{equation}
where the Jacobian $[A,B] := \z\cdot\nabla A\times \nabla B$ for
scalar fields $A,B$.  Here $\a\cdot\Dt{}\d{\x} = \a\cdot\d{\u} =
\a\cdot\partial_j\u\d{x}_j = a_j\nabla \bar u_j\cdot\d{\x}$ was
used along with Stokes theorem. The above is the statement of the
Kelvin circulation theorem for a general Lagrangian $\L[\u,h,\t]$.

Note that $\mathcal I$ is not conserved; it is created (or destroyed)
by the misalignment of the gradients of $\t$ and its dual
$h^{-1}\del{\L}{\t}$.  If $\partial D(\u)$ is replaced by $\partial
D$, then the Kelvin circulation is conserved because
$\smash{\oint_{\partial D} h^{-1}\del{\L}{\t} \nabla\t\cdot\d{\x}
= 0}$ by the assumed \emph{isopycnic} nature of the solid boundary
of the flow domain.

For the specific choice of Lagrangian for the IL$^0$PE \eqref{eq:L},
we have
\begin{equation}
   \mathcal I = \oint_{\partial D(\u)}(\u+\f)\cdot\d{\x} =
	\int_{D(\u)} h\bar q \d{}^2\x
\end{equation}
using Stokes theorem, where 
\begin{equation}
  \bar q = \frac{\z\cdot\curl\u + f}{h}
\end{equation}
is the potential vorticity. Noting that the rightmost equality in
\eqref{eq:KN} is $\smash{\int_{D(\u)} (2h)^{-1}[h,\t] \d{}^2\x}$,
by volume preservation we have
\begin{equation}
  \Dt{\bar q} = \frac{[h,\t]}{2h},
\end{equation}
i.e., potential vorticity is \emph{not} conserved.

\subsection{Lie--Poisson structure}

Consider the Hamiltonian
\begin{equation}
  \H[\m,h,\tt] := \int_D \m\cdot\u - \tfrac{1}{2}h|\u|^2 -
  h\u\cdot\f + \tfrac{1}{2}\tt h,
  \label{eq:H}
\end{equation}
which is nothing but the energy of the IL$^0$, given by
$\smash{\frac{1}{2}\int_D h|\u|^2 + \t h^2}$. One could guess it
or, much better,\citep{Holm-etal-21} obtain it via the Legendre
transform $(\u,h,\t)\mapsto (\m,h,\tt)$ defined by
\begin{equation}
  \H[\m,h,\tt] = \int_D \m\cdot\u - \L[\u,h,\t],
  \label{eq:lt}
\end{equation}
where (or upon realizing that)
\begin{equation}
  \m \big(= h(\u+\f)\big) = \del{\L}{\u},
\end{equation}
i.e., $\m$ is dual to $\u$.

Given the variational derivatives
\begin{equation}
  \del{\H}{\m} = \u,\quad
  \del{\H}{h} =  - \tfrac{1}{2}|\u|^2 - \u\cdot\f + \tfrac{1}{2}\tt,\quad
  \del{\H}{\tt} = \tfrac{1}{2}h,
\end{equation}
it is easy to guess that the Poisson tensor operator
should be given by
\begin{equation}
  \J := -
  \begin{pmatrix}
  \big((\cdot)\cdot\nabla\big)\m + \m\big(\nabla\cdot(\cdot)\big) +
  \big(\nabla(\cdot)\big)\cdot\m & h\nabla(\cdot) & \tt\nabla(\cdot) \\
  \nabla\cdot h(\cdot) & 0 & 0\\
  \nabla\cdot \tt(\cdot) & 0 & 0
  \end{pmatrix}
  \label{eq:J}
\end{equation}
in order for 
\begin{equation}
  \partial_t\p^i = \J^{ij}\del{\H}{\p^j},\quad \p := (\m,h,\tt),
\end{equation}
to give the IL$^0$PE (with the momentum equation written as
\eqref{eq:lp}, most naturally).  
\begin{remark}
  \citet{Holm-etal-21} give a stochastic version of the above $\J$
  in the variables $(\m,h,\t)$, which does not lead to a semidirect
  product Lie--Poisson bracket.  In turn, \citet{Dellar-03} gives
  a similar $\J$, but for an MHD system.  Actually, \citet{Dellar-03}
  never gives this $\J$!
\end{remark}
The above \emph{indeed} is a Poisson tensor operator since the
Poisson bracket,
\begin{equation}
  \{\U,\V\} := 
  \int_D \del{\U}{\p^i}\J^{ij}\del{\V}{\p^j},\quad
  \dot\U = \{\U,\H\},
\end{equation}
is skew-adjoint ($\{\U,\V\} = - \{\V,\U\}$) and satisfies the Jacobi
identity ($\{\U,\{\V,\W\}\} + \cp = 0$). It
is not difficult to show that\footnote{For this is better to
write the Poisson tensor as
\begin{equation}
  \J = -
  \begin{pmatrix}
  \bar m_j\partial_i + \partial_j\bar m_i & h\partial_i &
  \tt\partial_i \\
  \partial_jh & 0 & 0\\
  \partial_j\tt & 0 & 0
  \end{pmatrix}.
\end{equation}
}
\begin{equation}
  \{\U,\V\} \!=\! - \!\!\int_D\!\!
  \m\!\cdot\!\left[\left(\del{\U}{\m}\!\cdot\!\nabla\!\right)\del{\V}{\m}\! -\! 
  \left(\del{\V}{\m}\!\cdot\!\nabla\!\right)\del{\U}{\m}\right]
  \!+\! h\!\left(\del{\U}{\m}\!\cdot\!\nabla\del{\V}{h} \!-\!
  \del{\V}{\m}\!\cdot\!\nabla\del{\U}{h}\right) \!+\!
  h\leftrightarrow\tt
  \label{eq:PB}
\end{equation}
upon integrating by parts, assuming that 
\begin{equation}
  \del{\U}{\m}\cdot\n = 0 = \n\cdot \del{\V}{\m} \quad
  (\x\in\partial D),
  \label{eq:adm}
\end{equation}
which is the so-called \emph{admissibility condition}
\citep{McIntyre-Shepherd-87} for any functional of state.

\citet{Morrison-82} discusses general bracket forms and conditions
under which the Jacobi identity is satisfied; they note that
\eqref{eq:PB} is one such type of bracket.  An explicit proof, for
a bracket of the form \eqref{eq:PB} but including several densities,
is given in App.\ A.   The bracket \eqref{eq:PB} happens to be a
special case of the bracket  given by \citet{Morrison-Greene-80a}
for an MHD system when the magnetic field ($\vec{\mathrm B}$ in
their notation) is ignored; cf.\ their equation (9).  This connection
had remained elusive until now to the best of my knowledge.

\begin{remark}
  The first term in \eqref{eq:PB} is a Lie bracket for the Lie
  algebra of vector fields.  The other two terms arise from the
  extension of this Lie bracket to the semidirect product Lie algebra
  in which vector fields act separately on \emph{densities}.
\end{remark}

\begin{remark}
  The admissibility condition \eqref{eq:adm} does not guarantee
  that if $\U$ and $\V$ satisfy it, the result of the operation
  $\{\U,\V\}$ will also satisfy it.  Thus the Poisson bracket needs
  to be modified in the presence of a solid (or more generally free)
  boundary.  Work in this direction is presented in \citet{Lewis-etal-86},
  but more seems necessary.
\end{remark}

\subsubsection{Casimirs}

The quantity
\begin{equation}
  \C[\m,h,\tt]:= \int_D \big(\z\cdot\nabla\times h^{-1}\m\big)
  F(h^{-1}\tt) + hG(h^{-1}\tt)
  ,\quad \forall
  F,G(\,),
\end{equation}
is conserved as it can be directly verified noting that
$\z\cdot\nabla\times h^{-1}\m = h\bar q$, using $\smash{(\nabla\t)
F(\t) = \nabla\int^{\t} F(\t)\d{\t}}$ to be able to apply Stokes
theorem, and \emph{the boundary condition $\n\times\nabla\t = 0$
for $\x\in\partial D$}.

In order to be a Casimir, it must commute in the Poisson bracket
with any admissible functional or, equivalently,
$\smash{\J^{ij}\del{\C}{\p^j} = 0}$.  This is most easily done
in the original variables $(\u,h,\t)$ with respect to 
\begin{equation}
  \J := -
  \begin{pmatrix}
	 \bar q \z \times (\cdot) & \nabla(\cdot)  &
	 -(\cdot)h^{-1}\nabla\t\\
    \nabla\cdot (\cdot) & 0 & 0\\
    h^{-1}(\cdot)\cdot\nabla\t & 0 & 0
  \end{pmatrix}.
  \label{eq:Jpedro}
\end{equation}
The corresponding bracket, given by
\begin{equation}
  \{\U,\V\} \!=\! - \!\int_D\! q\z\cdot \del{\V}{\u}\times \del{\U}{\u}
  + \del{\U}{h}\nabla\cdot\del{\V}{\u}
  - \del{\V}{h}\nabla\cdot\del{\U}{\u}
  -h^{-1}\nabla\t\cdot\left(\del{\U}{\t}\del{\V}{\u} -
  \del{\V}{\t}\del{\U}{\u}\right),
  \label{eq:PBpedro}
\end{equation}
was shown by \citet{Ripa-GAFD-93} to satisfy the Jacobi identity
while is not Lie--Poisson. In fact, the bracket \eqref{eq:PB}
follows from the above bracket under the
transformation (chain rule)
\begin{equation}
  \left.\del{}{h}\right\vert_{\u,\t} =
  \left.\del{}{h}\right\vert_{\m,\tt} +
  h^{-1}\m\cdot\del{}{\m} + \tt h^{-1}\del{}{\tt}
\end{equation}
and similarly for the other variables. The $\J$ given in \eqref{eq:Jpedro}
gives the IL$^0$PE with the momentum equation \eqref{eq:mom} directly
in the form
\begin{equation}
  \partial_t\u + h\bar q \z \times \u + \nabla\big(\t h +
  \tfrac{1}{2}|\u|^2\big) - \tfrac{1}{2}h\nabla\t = 0.
\end{equation}
Then one can write the Casimir (as it was originally proposed
by \citet{Ripa-GAFD-93}) as
\begin{equation}
  \C[\u,h,\t]:= \int_D h \big(qF(\t) + G(\t)\big),
\end{equation}
(note that it is $\t$, rather than $\tt = h\t$) whose functional
derivatives are\footnote{The boundary term $\smash{F(\t)\vert_{\partial
D}\oint_{\partial D} \delta\u\cdot\d{\x}}$ vanishes identically
since the circulation along the \emph{solid} boundary $\partial D$
is constant; cf.\ Sec.\ \ref{sec:KN}.}
\begin{align}
  \del{\C}{\u} &= - \z\times\nabla F(\t),\\
  \del{\C}{h} &=  G(\t),\\
  \del{\C}{\t} &= h \big(qF'(\t) + G'(\t)\big).
\end{align}
Then one readily verifies that $\smash{\J^{1j}} = -q\nabla F(\t) -
\nabla G(\t) + q\nabla F(\t) + \nabla G(\t) \equiv 0$, $\smash{\J^{2j}}
= \nabla\cdot \z\times\nabla F(\t) \equiv 0$, and $\smash{\J^{3j}}
= -h^{-1} F'(\t) \z\cdot \nabla\t \times \nabla\t \equiv 0$.

The Poisson bracket \eqref{eq:PBpedro} (given in \citet{Ripa-GAFD-93})
happens to be the bracket (6) in \citet{Morrison-Greene-80a} with
no magnetic field ($\vec{\mathrm B} = 0$). To explicitly see the
emergence of the Poisson tensor \eqref{eq:Jpedro} from Morrison and
Green's \cite{Morrison-Greene-80a} ``hydrodynamics equations,'' one
must note that the pressure gradient force of that set
$\rho^{-1}\nabla(\rho^2 U_\rho) = \nabla(\rho U_\rho + U) - U_s\nabla
s$ since $U$ is a function of $(\rho,s)$ where $\rho \leftrightarrow
h$ and $s \leftrightarrow \t$. Their set does not include Coriolis
force, and the pressure gradient force (in our notation) $\nabla(U
+ h\partial_h U) - \partial_{\t}U\nabla\t$, for some $U(h,\t)$,
instead of $\smash{\nabla(h\t) - \frac{1}{2}h\nabla\t}$.  In other
words, the specific set of dynamical equations depend on the specific
choice of Hamiltonian, which in Morrison and Green's
\cite{Morrison-Greene-80a} case took to form $\smash{\H =
\int_D\frac{1}{2}h|\u|^2 + hU(h,\t)}$.  The geometry of the system,
independent of its specific form, is determined by the Poisson
bracket.  Of course, the choice $\smash{U = \frac{1}{2}h\t}$ gives
the IL$^0$PE's pressure gradient force.

\section{Discussion and outlook}

With the last comment in mind, it turns out that the IL$^0$ can be
generalized to include a pressure gradient force of the form
$h^{-1}\nabla h^2\partial_h\varphi(h,\t)$ where $\varphi(\,,\hspace{.1em})$
is arbitrary.  The choice $\varphi = \tfrac{1}{2}h\t$ gives the
IL$^0$ pressure force, as noted.  The implications of such a
generalization await to be assessed.  Moreover, a further generalization
of the IL$^0$ model is possible representing an additional extension
of the semidirect product Lie algebra bracket \eqref{eq:PB} to
include an arbitrary number of density forms.  A particular choice
gives a model, which I will call IL$^{0.5}$, that has buoyancy also
varying (linearly) in the vertical direction.  This will enable one
to model important processes that lie beyond the scope of the IL$^0$
class like mixed-layer restratification resulting from baroclinic
instability.\citep{Boccaletti-etal-07} Investigating the geometric
properties of the IL$^{0.5}$, which is similar to a model used
earlier in equatorial dynamics,\citep{Schopf-Cane-83} and its
quasigeostrophic version including their performance in direct
numerical simulations is the subject of ongoing work.

\paragraph{Acknowledgements.} 

I thank Prof.\ Philip J.\ Morrison for suggesting the connection
with \citet{Morrison-Greene-80a} and for stimulating discussions
on Poisson brackets during the Aspen Center for Physics workshop
``Transport and Mixing of Tracers in Geophysics and Astrophysics,''
where these notes were written and additional ongoing work was
initiated.

\appendix

\section{Jacobi identity}

Let 
\begin{equation}
  \{\U,\V\}^{\m} := \int_D \m\cdot [\U_{\m},\V_{\m}],\quad
  \{\U,\V\}^{\rho_\alpha} := \int_D
  \rho_\alpha \big(\U_{\m}\cdot\nabla\V_{\rho_\alpha} -
  \V_{\m}\cdot\nabla\U_{\rho_\alpha}\big).
\end{equation}
Here the shorthand notation $\U_\mu := \del{\U}{\mu}$ was used
along with the \emph{commutator} 
\begin{equation}
  [\a,\b] := (\a\cdot\nabla)\b - (\b\cdot\nabla)\a,
  \label{eq:com}
\end{equation}
which is antisymmetric, i.e.,
\begin{equation}
  [\a,\b] = - [\b,\a]
\end{equation} 
and satisfies the Jacobi identity, viz., 
\begin{equation}
  [[\a,\b],\c] + \cp = 0
  \label{eq:com-jac}
\end{equation}
The first property is obvious; the second one involves some algebra
but is otherwise quite straightforward to verify:
\begin{align}
  [[\a,\b],\c] = {} &
  + 
  \underbrace{\Big(\big((\a\cdot\nabla)\b\big)\cdot\nabla\Big)\c}_{\boxed{1}}
  - 
  \underbrace{\Big(\big((\b\cdot\nabla)\a\big)\cdot\nabla\Big)\c}_{\boxed{2}}\nonumber\\
  &
  - 
  \underbrace{(\c\cdot\nabla)\big((\a\cdot\nabla)\b\big)}_{\boxed{3} 
  + (\c\a:\nabla\nabla)\b}
  +
  \underbrace{(\c\cdot\nabla)\big((\b\cdot\nabla)\a\big)}_{\boxed{4}
  + (\c\b:\nabla\nabla)\a}\label{eq:com-jac-1}\\
  [[\c,\a],\b] = {} &
  + 
  \underbrace{\Big(\big((\c\cdot\nabla)\a\big)\cdot\nabla\Big)\b}_{\boxed{3}}
  - 
  \underbrace{\Big(\big((\a\cdot\nabla)\c\big)\cdot\nabla\Big)\b}_{\boxed{5}}\nonumber\\
  &
  - 
  \underbrace{(\b\cdot\nabla)\big((\c\cdot\nabla)\a\big)}_{\boxed{6}
  + (\b\c:\nabla\nabla)\a}
  +
  \underbrace{(\b\cdot\nabla)\big((\a\cdot\nabla)\c\big)}_{\boxed{2}
  + (\b\a:\nabla\nabla)\c}\label{eq:com-jac-2}\\
  [[\b,\c],\a] = {} &
  + 
  \underbrace{\Big(\big((\b\cdot\nabla)\c\big)\cdot\nabla\Big)\a}_{\boxed{6}}
  - 
  \underbrace{\Big(\big((\c\cdot\nabla)\b\big)\cdot\nabla\Big)\a}_{\boxed{4}}\nonumber\\
  &
  - 
 \underbrace{(\a\cdot\nabla)\big((\b\cdot\nabla)\c\big)}_{\boxed{1}
 + (\a\b:\nabla\nabla)\c}
  +
  \underbrace{(\a\cdot\nabla)\big((\c\cdot\nabla)\b\big)}_{\boxed{5}
  + (\a\c:\nabla\nabla)\b},\label{eq:com-jac-3}
\end{align}
where $\a\b : \c\mathbf d = a_ib_j c_id_j$. Adding
\eqref{eq:com-jac-1}--\eqref{eq:com-jac-3} with $\a\b = \b\a$ in
mind proves \eqref{eq:com-jac}.$\square$

Now,
\begin{equation}
  \{\U,\V\}^{\m} = - \{\V,\U\}^{\m},\quad
  \{\U,\V\}^{\rho_\alpha} = - \{\V,\U\}^{\rho_\alpha},
\end{equation}
manifestly. Our goal is the demonstrate that
\begin{equation}
  \{\U,\V\} := \{\U,\V\}^{\m} +
  \sum_\alpha\{\U,\V\}^{\rho_\alpha}  
  \label{eq:PBalpha}
\end{equation}
satisfies $\{\{\U,\V\},\W\} + \cp = 0$. The interest is in $\rho_1
= h$ and $\rho_2 = \tt$, but the argument can be extended to include
an arbitrary number of \emph{densities}. More precisely, we seek
to show that
\begin{equation}
  \{\{\U,\V\},\W\} = \{\{\U,\V\},\W\}^{\m} + \sum_\alpha
  \{\{\U,\V\},\W\}^{\rho_\alpha}
  \label{eq:mrho}
\end{equation}
vanishes upon $\cp$. To do it, we consider each term in
\eqref{eq:mrho} at a time, with the following in mind:
\begin{equation}
  \{\U,\V\}_{\m} = [\U_{\m},\V_{\m}],\quad
  \{\U,\V\}_{\rho_\alpha} = \W_{\m}\cdot\nabla \big(\U_{\m}\cdot\nabla\V_{\rho_\alpha} -
  \V_{\m}\cdot\nabla\U_{\rho_\alpha}\big).
  \label{eq:der}
\end{equation}
Indeed, $\{\U,\V\}[\m,\rho_1,\rho_2,\cdots] = \{\U,\V\}^{\m}[\m] +
\{\U,\V\}^{\rho_1}[\rho_1] + \{\U,\V\}^{\rho_2}[\rho_2] + \cdots$,
with the functional dependence of each term on the argument being
linear.

Let us start with the $\m$ bracket, which, using the left relationship
in \eqref{eq:der}, reads
\begin{align}
  \{\{\U,\V\},\W\}^{\m} &= - \int_D\m\cdot
  \left[\{\U,\V\}_{\m},\W_{\m}\right] = - \int_D\m\cdot
  \left[[\U_{\m},\V_{\m}],\W_{\m}\right].
\end{align}
Since $[\,,\hspace{.1em}]$ satisfies the Jacobi identity, we readily find
\begin{equation}
  \{\{\U,\V\},\W\}^{\m} + \cp = 0.
  \label{eq:jac_m}
\end{equation}

We now turn to the $\rho_\alpha$ brackets in \eqref{eq:mrho}, which
require more elaboration. It is enough to consider one term only,
though. More precisely,
\begin{align}
  \{\{\U,\V\},\W\}^{\rho_\alpha} = {} & - \int_D
  \rho_\alpha\left(\{\U,\V\}_{\m}\cdot\nabla\W_{\rho_\alpha} -
  \W_{\m}\cdot\nabla\{\U,\V\}_{\rho_\alpha}\right)\nonumber\\ =
  {} & - \int_D \rho_\alpha[\U_{\m},\V_{\m}]\cdot \nabla\W_{\rho_\alpha}
  - \rho_\alpha\W_{\m}\cdot\nabla \big(\U_{\m}\cdot\nabla\V_{\rho_\alpha} -
  \V_{\m}\cdot\nabla\U_{\rho_\alpha}\big)\nonumber\\
  = {} & - \int_D \rho_\alpha\big([\U_{\m},\V_{\m}]\big)\cdot \nabla\W_{\rho_\alpha}
  \nonumber \\ & + \int_D \rho_\alpha\Big(\big((\W_{\m}\cdot\nabla)\U_{\m}\big)\cdot\nabla\V_{\rho_\alpha} -
  (\W_{\m}\cdot\nabla)\V_{\m}\big)\cdot\nabla\U_{\rho_\alpha}\Big)
\end{align}
where, in order, we took into account \eqref{eq:der} and
\begin{equation}
  (\a\cdot\nabla)\b\cdot\mathbf c = ((\a\cdot\nabla)\b)\cdot \mathbf
  c + \a\b:\nabla\mathbf c
\end{equation}
(recalling that $\a\b = \b\a$). More explicitly, omitting the
$-\int_D\rho_\alpha$, we have
\begin{align}
  \{\{\U,\V\},\W\}^{\rho_\alpha} = {} &
  + 
  \underbrace{\big((\Fm\cdot\nabla)\Gm\big)\cdot\nabla\Hs}_{\boxed{1}}
  - 
  \underbrace{\big((\Gm\cdot\nabla)\Fm\big)\cdot\nabla\Hs}_{\boxed{2}}\nonumber\\
  &
  - 
  \underbrace{\big((\Hm\cdot\nabla)\Fm\big)\cdot\nabla\Gs}_{\boxed{3}}
  +
  \underbrace{\big((\Hm\cdot\nabla)\Gm\big)\cdot\nabla\Fs}_{\boxed{4}}\label{eq:rho-jac-1}\\
  \{\{\W,\U\},\V\}^{\rho_\alpha} = {} &
  + 
  \underbrace{\big((\Hm\cdot\nabla)\Fm\big)\cdot\nabla\Gs}_{\boxed{3}}
  - 
  \underbrace{\big((\Fm\cdot\nabla)\Hm\big)\cdot\nabla\Gs}_{\boxed{5}}\nonumber\\
  &
  - 
  \underbrace{\big((\Gm\cdot\nabla)\Hm\big)\cdot\nabla\Fs}_{\boxed{6}}
  +
  \underbrace{\big((\Gm\cdot\nabla)\Fm\big)\cdot\nabla\Hs}_{\boxed{2}}\label{eq:rho-jac-2}\\
  \{\{\V,\W\},\U\}^{\rho_\alpha} = {} &
  + 
  \underbrace{\big((\Gm\cdot\nabla)\Hm\big)\cdot\nabla\Fs}_{\boxed{6}}
  - 
  \underbrace{\big((\Hm\cdot\nabla)\Gm\big)\cdot\nabla\Fs}_{\boxed{4}}\nonumber\\
  &
  - 
 \underbrace{\big((\Fm\cdot\nabla)\Gm\big)\cdot\nabla\Hs}_{\boxed{1}}
  +
 \underbrace{\big((\Fm\cdot\nabla)\Hm\big)\cdot\nabla\Gs}_{\boxed{5}}\label{eq:rho-jac-3}.
\end{align}
Adding \eqref{eq:rho-jac-1}--\eqref{eq:rho-jac-3}, one obtains
\begin{equation}
  \{\{\U,\V\},\W\}^{\rho_\alpha} + \cp = 0.
  \label{eq:jac_rho}
\end{equation}

Thus, \eqref{eq:jac_m} and \eqref{eq:jac_rho} together
produce the desired result:
\begin{equation}
  \{\{\U,\V\},\W\} + \cp = 0.\,\blacksquare
\end{equation}

We close by indicating in Table \ref{tab:cas} the Casimirs of the
bracket \eqref{eq:PBalpha}.

\begin{table}
  \centering
  \renewcommand*{\arraystretch}{1.5}
  \begin{tabular}{cc}
	 \hline\hline
	 $1 < \alpha \le$ & $\C = \int_D $\\
	 $1$ & $\rho_1F(\rho^{-1}_1\z\cdot\nabla\times\rho^{-1}_1\m)$\\
	 $2$ & $(\z\cdot\nabla\times\rho^{-1}_1\m) F(\rho_2) + \rho_1  G(\rho_2)$\\
	 $n \ge 3$ & $\rho_1 F(\rho_2,\dotsc,\rho_n)$\\
	 \hline
  \end{tabular}
  \renewcommand*{\arraystretch}{0.5}
  \caption{Casimirs of the Poisson bracket \eqref{eq:PBalpha}.}
  \label{tab:cas}
\end{table}

\bibliographystyle{mybst}
\bibliography{fot}

\end{document}